\documentclass[10pt,letterpaper]{article}
\usepackage[top=0.85in,left=2.75in,footskip=0.75in]{geometry}

\usepackage{amsmath,amssymb}

\usepackage{changepage}

\usepackage[utf8]{inputenc}

\usepackage{textcomp,marvosym}

\usepackage{cite}

\usepackage{nameref,hyperref}

\usepackage[right]{lineno}

\usepackage{microtype}
\DisableLigatures[f]{encoding = *, family = * }

\usepackage[table]{xcolor}

\usepackage{array}

\newcolumntype{+}{!{\vrule width 2pt}}

\newlength\savedwidth

\raggedright
\usepackage[aboveskip=1pt,labelfont=bf,labelsep=period,justification=raggedright,singlelinecheck=off]{caption}

\makeatletter
\renewcommand{\@biblabel}[1]{\quad#1.}
\makeatother

\usepackage{lastpage,fancyhdr,graphicx}
\usepackage{epstopdf}
\fancyheadoffset[L]{2.25in}
\fancyfootoffset[L]{2.25in}
 \date{}

\begin{document}
\vspace*{0.2in}

\begin{flushleft}
{\Large
\textbf\newline{Augmenting reality:\\ On the shared history of perceptual illusion and video projection mapping} 
}
\newline
\\
Alvaro Pastor
\\
\bigskip
Computer Science, Multimedia and Telecommunications Department. \\Universitat Oberta de Catalunya. Barcelona, Spain.
\\
\bigskip

alvaropastor@uoc.edu

\end{flushleft}

\section*{Abstract}
Perceptual illusions based on the spatial correspondence between objects and displayed images have been pursued by artists and scientists since the 15th century, mastering optics to create crucial techniques as the linear perspective and devices as the \textit{Magic Lantern}. Contemporary video projection mapping inherits and further extends this drive to produce perceptual illusions in space by incorporating the required real time capabilities for dynamically superposing the imaginary onto physical objects under fluid real world conditions. A critical milestone has been reached in the creation of the technical possibilities for all encompassing, untethered synthetic reality experiences available to the plain senses, where every surface may act as a screen and the relation to everyday objects is open to  perceptual alterations.


\section*{Introduction}
The advancement and popularization of computer technologies over the past three decades has had a significant effect not only on the way images are created, but also on the techniques and technologies which enable their presentation. One of the most popular emerging practices is video projection mapping, also referred as video mapping or 3D mapping, a terminology which stands for a set of imaging techniques in service of the spatially coherent projection of bi dimensional images onto physical three dimensional objects~\cite{yun,naimark1}.
Despite its seemingly novelty in its use in the fields of communication, advertising and visual arts in the recent years, the current practice of video projection mapping is the result of several centuries of development and perfecting of a vast set of techniques derived from optics and the sciences of human perception, as well as mathematical operations for image transformations, and the evolution projection devices. Across all these periods in history, the production of perceptual illusions remains as the fundamental shared goal, of which important milestones are the invention of linear perspective techniques and its application in \textit{Trompe l'oeil} and \textit{Quadratura} installations during the Renassaince, the immersive theatrical qualities of \textit{Panorama} and \textit{Phantasmagoria} in the 18th century, the development of real time 3D computer rendering capabilities and the popularization of video projection mapping during the last part of the 20th century.
The present article analyzes the evolution of video projection mapping in relation to the history of perceptual illusion practices, considering common design methodologies, tools, and implemented techniques, in a transversal selection of scientific and artistic practices, spreading from static compositions on a canvas, to dynamic substitutions in three dimensional space. A commentary on the emerging perceptual and cognitive phenomena as a result of the spatial distribution of synthetic contents is provided, presenting the notion that the current stance of real time dynamic forms of video projection mapping constitutes a turning point in the quest for replacing the real world with synthetic versions that would offer unprecedented means for visual identity and meaning transformations beyond material constraints.

\section*{Spatial illusion in the Renassaince}
Before the 14th Century few attempts were made to create the illusion of three dimensional space and to realistically depict the spatial world in art. The Italian masters Giotto and Duccio during the late thirteenth and early fourteenth centuries, began to explore the idea of depth and volume in their artwork, and may be credited with introducing an early form of three dimensional painting, using shadows and contrasts in colors to create the perceived illusion of depth~\cite{giotto,schet}. Although it was still very far from the accurate and robust three dimensional representations popular in imagery today, this technique known as \textit{Chiaroscuro} helped artists to create the illusion of volume and depth. Popularized during the beginning of the fifteenth century, Chiaroscuro was mastered especially for the processes of modeling and representing the human body and was extensively used during the sixteenth and seventeenth centuries in the Mannerism and Baroque eras as one of the standards for approximating three dimensional representation.
Not satisfied with the state of the art, Florentine painters of the early fifteenth century, wanted to develop painting as a science derived from Euclidean geometry influenced by the Neoplatonic intellectual postulates. In this sense, they did not conform with the intuitive Chiaroscuro technique, but sought scientific methods of representing reality based on mathematical laws, thus giving birth to the knowledge that ultimately led to the devise of the linear perspective system. The linear perspective system projected the illusion of depth onto a two dimensional plane by use of vanishing points to which all lines converged at eye level on the horizon, accurately reproducing the way that objects appear smaller when they are further away, and the way parallel lines appear to meet each other at a point in the distance.

Between 1415 and 1420, the Italian architect Filippo Brunelleschi, considered one of the founding figures of the Renassaince, conducted and documented a series of experiments which helped him develop the correct perspective techniques for accurately representing three dimensional volumes on a flat canvas. One of its results portrayed in accurate spatial proportion and organization, the Florence Baptistery and the Palazzo Vecchio building seen from the front gate of the unfinished cathedral, affording to the spectator the full illusion of being present in the scene. These technical advancements where later systematized by Leon Battista Alberti in his 1436 work \textit{De Pictura}, one of the most important treatises on painting of the Renaissance, enabling artists for the first time to paint imaginary scenes with perfectly accurate three dimensional realism by following the set of rules of perspective studied by Brunelleschi~\cite{albert, alberti2,dunning, brock15}. Gradually, a standardized model of human visual perception was consolidated, from which a set of mathematical rules was derived and applied to govern the disposition of pictorial elements in order to produce a credible illusion of three dimensionality. \\
For the next five centuries, Brunelleschi’s system of perspective and its derived techniques were used to create the illusion of depth on the picture plane, and served as the basis of some of the great art of Western culture~\cite{gablik,gombrich, verstegen}. Among the earliest of those, The Brera Madonna created between 1472 and 1474 by the Italian artist Piero della Francesca. It is a large painting depicting a group of religious figures and icons at a fictitious apse of the church, meticulously rendered using the innovations of linear perspective techniques. But not only did perspective techniques allowed to produce a coherent illusion of space on a flat surface, but perspective was also used to organize and highlight particular elements of the composition, thus allocating meaning in relation to specific spatial positions in the canvas~\cite{banker, shw}. For example, the geometric center of the face of the main character coincides with the vanishing point of the perspective system used in the scene, as well as the egg which hangs from a shell - shaped element below the apse, is perfectly aligned with the scene's vanishing point and marks a perpendicular line to the scene's horizon.
Tommaso di Ser Giovanni di Simone, best known as \textit{Masaccio} and regarded as the first great Italian painter of the \textit{Quattrocento} period of the Italian Renaissance, painted \textit{The Holy Trinity} in 1425 in the dominican church of Santa Maria Novella in Florence. This pioneering mural painting not only accurately presented the illusion of three dimensions using linear perspective techniques, but also engaged in a coherent composition with the pre existing architectural elements that surrounded the mural, mixing three dimensional painted and architectural elements to enhance the illusory effect. This specific type of illusion supported by the coherent merger between painting techniques and architectural elements was known as Quadratura~\cite{masa}.
Artist Andrea Mantegna further experimented developing illusionistic ceiling paintings using perspective techniques to create the perceptual illusions and was responsible in 1465 for one of the most significant examples of Quadratura. Located in the Camera degli Sposi in the Ducal Palace of Mantua, this Trompe l'oeil was applied to ceiling paintings to be seen by spectators from below and upward and presented the illusionistic painting of an oculus that opens to the sky and a number of fictional characters who look down towards the spectators~\cite{wade}. Performed by Andrea Pozzo in 1688, \textit{The Apotheosis of St Ignatius} in Sant Ignazio in Rome is another remarkable example of Quadratura, presenting in a 55 feet wide ceiling a composition combining fictional characters and architectural elements. A new type of site specific art form began maturing, unifying architecture, painting and sculpture, and being  conveyed to spectators by means of a full body spatial experience. In a sense, this almost theatrical experience was prescient of the aesthetic ideals of \textit{Gesamtkunstwerk}, postulated years later in mid nineteenth century by the German opera composer Richard Wagner.

\section*{Panorama and Phantasmagoria}
Baroque painters from seventeenth century, notably Caravaggio, Bernini, Rubens, Rembrandt, Velazquez and Vermeer, used the now long established three dimensional representation system in hundreds of varied approaches, as did the painters from the Neoclassicist movement, of whom a notable example is the work of Giovanni Batista Piranesi. \textit{Carceri d'invenzione}, a series prints produced between 1745 and 1761 by Piranesi depicting somber fictional scenes full of mysterious characters, bridges, stairs and mechanisms, trough meticulous utilization of the perspective technique. However, few artists during these decades shared the interest for the experimentation and development of perceptual illusions via spatial experience. In this sense, a major referent is the work of Robert Barker, an Irish painter that developed the idea of erecting a circular plan building to house near real scale paintings featuring panoramic landscapes. In 1792 the Panorama was born, supported on the knowledge on perspective and color contrast practiced during the last two centuries, its paintings covered 360 degrees around the spectator in the horizontal plane and reached 15 meter of height.
Soon, Barker's Panorama grew in size and complexity, and the circular plan installation incorporated topographies and architectural and elements intended to match the painted scene thus increasing its illusory effect in a similar way as Quadratura incorporated architectural elements to its composition. This way, flooring, stairs, railings, water fountains or other urban furniture, not only to regulate the displacements of spectators inside the building, but more importantly, to serve as immersive cues, formal and semantic links between the depicted scenes and the notion of what was accountable as real world experience~\cite{panorama, flieth}. The Panorama continued its popularization in most of Western Europe and North America as an entertainment form, successfully hosting the representation of famous historical battles for the great public. In a time when few people could afford to travel, the Panorama allowed to walk through distant landscapes and contemplate historical events from privileged points of view, pioneering in a primitive form of narrative distributed in space and time, in a sense similar as it is now possible to explore in a virtual reality experience or a video game~\cite{finn,finn2}.
During the precedent decades, the rapid increase of knowledge in optics was not only useful for development and use of perspective related techniques, but also supported a number of relevant inventions regarding optical instruments, among those, light reflection and projection devices. The Jesuit Athanasius Kircher in his book \textit{Ars Magna Lucis et Umbrae} from 1645, described a primitive projection system with a focusing lens and a concave mirror reflecting sunlight that served as canvas to display text or pictures known as the \textit{Steganographic Mirror}. Further elaborating, dutch scientist Christiaan Huygens in 1659 documented the invention of the \textit{Magic Lantern}, an early type of image projector that used semi transparent, hand painted glass plates, one or more lenses, and a light source~\cite{swan}. The Magic Lantern was then consolidated, and increasingly used for education and entertainment purposes during the 18th century~\cite{kember}. In this period, although the basic design of the Magic Lantern suffered various modifications in relation to the light source capabilities and the characteristics of lenses in use, its main fundamental concept remain unchanged as well as its basic operation. 
By the end of 18th century, Belgian Physicist and cleric Ettiene Gaspar Robert, and one of the first lanternists of all time Paul Philidor, created a new form of illusion performance in the form of a Magic Lantern show known as Phantasmagoria, a theatrical show that consisted of a mixture of frontal and retro projection of images, from simultaneous magic lanterns, on canvases and solid elements in front of the sitting public~\cite{hecht, barber, phanta}. During the 19th century, the invention was further sophisticated, with more powerful light sources, incorporation of photographic impressions onto glass plates, improvements in optics, and mechanisms that allow the projection of sequences of images, helping consolidate the expressive maturity of a new type visual spectacle still greatly influenced by the use of perspective and contrast in colors to achieve volumetric realism~\cite{clarke}.  Following this theatrical illusory tradition, Louis Daguerre in 1822 developed the \textit{Diorama} for a light based type of scene representation device, that featured two immense paintings and various interchangeable moving elements, such as human figures, animals or fictional characters, lit from the front and through the back inside an otherwise pitch black, rotating auditorium. The Diorama combined techniques of opaque and translucent painting and manipulating light in a live spectacle, with color contrast and perspective techniques, to produce a plausible scene with accurate spatial representation. In a large scale version, the Diorama frame could reach 10 meter wide and feature interposing translucent elements of different opacities and colors, actuated by means of ropes that affect the color tone of selective parts of the frame, for example, to simulate intense fog or the shining sun~\cite{Hankins}. 
A few decades later, the German theater director and producer Erwin Piscator developed the most consistent scenography experiments using the technological possibilities of the early 20th century to support epic theater works emphasizing the socio political content of drama. Notably, his early production \textit{St\"urmflut} from 1926 the stage was composed by a translucent background screen in a solid black frame with variable aperture serving as background displaying supporting images from four retro projectors for performing artists.

\section*{Performatic spacetime}
Although most of the works of Impressionist painters such as Renoir, Sisley, Monet and Pissarro and Post Impressionists such as Van Gogh, Gauguin and Seurat also depended on these mimetic spatial representation techniques, a significant transformation took place in the broad scientific and artistic domains at the end of the nineteenth century. It challenged the absolute centralized perspective standard, in which the images orbited around the static spectator's eye, increasingly giving way to the development of the multiple viewpoints systems, which aimed to capture time lapses in which the scene should be considered not as static but in motion.
The idea of multiplicity in viewpoints was mainly developed in the arts by the Cubist movement, fundamentally Picasso and Braque in their invention of techniques such as faceting, passage and multiple perspective. This type of representations of multiple viewpoints in space and time, fused in one single image, consolidated as one of the movement's most acclaimed traits. Importantly, these innovations may be interpreted as the movements' critical philosophical stance regarding perspective and the capacity of human beings to faithfully grasp and represent reality. For the Cubists, art would not serve to be a mimesis of reality, as an objective mirror snapshot of \textit{what is there}, but would serve to represent things as experienced in space and time, a subjective snapshot of \textit{what is sensed}. This new method aimed to account for change and flux in a scene, and create a performative situation in which the interpretative role of the spectator becomes central to the evocative power of the Cubist's images~\cite{edgerton}.
Following these lines, Albert Gleizes and Jean Metzinger in the 1912 book \textit{Du "Cubisme"}, the first major text on Cubism, explicitly related the pursue of multiple perspective to an attempt to understand time as a continuum. This may be interpreted as giving symbolic expression to the notion of duration proposed by the philosopher Henri Bergson, according to which, life is subjectively experienced as a continuum, with the past flowing into the present and the present merging into the future~\cite{cubisme,bergson}. Early Futurist paintings by Umberto Boccioni, Giacomo Balla or Gino Severini where also highly influenced by this search for multiple perspective and simultaneity. This technique of representing multiple viewpoints in space and time is advanced in complexity and scale in Gleizes' \textit{Le Dépiquage des Moissons}, displayed at the 1912 \textit{Salon de la Section d'Or}, and Robert Delaunay's City of Paris, exhibited at the \textit{Indépendants} in 1912. 
As the invention of linear perspective in fifteenth century Renaissance was a major manifestation of the establishment of the strong anthropocentric interpretation of the world, in which the whole of its Science and Philosophy was founded upon, the multiple perspective paradigm that arrived at the dawn of the 20th century marked the ultimate disenchantment of human as the measure of all things, welcoming novel ideas from Physics regarding the nature of space and time, as well as the atom and atomic dynamics, which ultimately favored many of the ideas that shaped the following decades: The relative over the absolute, multiplicity over centralization, dynamic co creation over pre determination~\cite{lepenies}. Whereas Renaissance artists strove to represent the real by offering viewers an archetypical perceptual illusion, Cubist illusionism subverts this kind of realistic representation by calling attention to its own artifice, to its own perspectival manipulations, and thus to the problematic nature of representation and illusion.
\section*{Expanded cinema}
During the 1950's The Czech artist Josef Svoboda pursued a type of non verbal theatrical format where actors perform in harmony along with moving images projected by a set of Magic Lanterns and elements of the scenography that supported the performative meaning production. This poetic synthesis of traditional methods with technical innovation including architecture, dance, cinema and theater, that in a sense revived Richard Wagner's \textit{Gesamtkunstwerk}, secured Czechoslovakia's best score in the Brussels World Exhibition of 1958. Svoboda not only mastered the classical perspective and color contrast techniques to put adequate spatial illusion in action on the stage, but also multiplied its evocative power by profiting from the emerging interactions between the characteristics of the selected physical elements, the performing actors and the content of the projected images in the composition~\cite{burian}. Moreover, part of this new theatrical format was Svoboda's \textit{Polyekran}, a multi projection system, consisting of screens of various shapes and dimensions suspended with steel cables in arbitrary angulations on a black background. Eight projectors of slides and seven movie projectors, along with synchronized stereo helped completing this instrument in service of the total work of art.\\
The immersive quality of spatially distributed images was also among the interests of the experimental filmmaker Stan VanDerBeek , who between 1963 and 1965 developed his \textit{Movie Drome}, a type of semi spherical theater where people would lie down and experience movies all around them. Notably, influenced by Buckminster Fuller’s dome constructions and Robert Barker's Panorama public installations, VanDerBeek envisioned the Movie Drome as a prototype for a global two way communication system consisting of a network of Movie Dromes linked to orbiting satellites that would store and transmit images between them. Although far from fully accomplishing it vision, Movie Dromes did put in place a new set of challenges for the traditional use of perspective and linear narratives to represent reality. Instead, VanDerBeek presented a multiple perspective design, that achieved multiple gazes over multiple simultaneous events, resulting in a fragmented non linear narrative, a spatially distributed moving image collage~\cite{sutton,ferguson}. Distinct from traditional cinema, this new type of experience sometimes characterized as \textit{Expanded Cinema}~\cite{youngblood} had a clear emphases on breaking the cinematography from its linear and concentrated constraints to give way to a non linear, distributed in space, multi sensory experience, aimed at a type of audience engagement that anticipates contemporary art's interactivity and participatory practices.
Audience engagement produced by carefully controlled perceptual illusions was also researched for popular entertainment purposes. As a proof, on 9 august 1969 the doors to the \textit{Haunted Mansion} spectacle at Disneyland opened to the public, featuring a number of light based perceptual illusions placed inside a three story Victorian house structure built in the New Orleans Square in Disney's Adventure Park in California. The project gathered the works of artists and storytellers Yale Gracey, Marc Davis, Claude Coats, Xavier Atencio and among those, Rolly Crump's innovative \textit{Singing Busts} installation~\cite{disney}. This work consisted of five singing busts, brougth to life by the \textit{Grim Grinning Ghosts} actors, pre recorded in 16mm film singing the theme song of the ride, and then projected onto the plaster busts modeled after each character. By controlling the lighting and contrast of the scene, together with the predetermined path and vantage point of user's, the designers could obtain an unprecedented multi sensory illusion that may be considered as direct precursor of current virtual and augmented reality paradigms.\\
Since the mid 1980s, numerous examples exist of re enactment of cinematographic scenes by the use of video sequences in combination with immersive objects and architectures. Among those the works of Tony Oursler who as one of the key figures in the development of video art experimented early on with the moving image that extended beyond the borders of the TV monitor~\cite{kaye}. Oursler used projections on sculptures and specially crafted scenographic installations. His early 1990s dolls and dummy works or \textit{Troubler} from 1996, rely entirely in the spatial and semantic correspondence between projected film and dolls which are strongly anthropomorphized~\cite{london}. Using centuries old knowledge on optics and perception, controlling light and shadow and the spectator's point of view Oursler mastered a unique type of cinematographic composition based on the combination of found objects, life-like moving faces and archetypal elements coming into life by virtue of the superimposed projected narratives~\cite{malt}.\\
Michael Naimark, is one key multimedia artist and researcher who has also explored spatialized representation. His \textit{Displacements} intallation, developed from 1980 to 1984, further advanced the spatial and semantic interactions between images and surfaces, and is considered among the most important seminal works in this domain. An immersive installation that projected a pre recorded 16mm film on an archetypal living room completely covered in white paint, matching the spatial location of the projected images with selected objects of the living room. The projector located on a rotating turntable at the room's center slowly directed the spectators' gaze towards the unwrapping narrative, with the appearance of the characters and objects as three dimensional. 
Artist Jeffrey Shaw has also undertaken since the late 1960s, research on types of Expanded Cinema and the development of various innovative spatial illusion strategies. For example \textit{Corpocinema} from 1967, involved projection into a dome, and the {Diadrama} from 1974 was projected onto a 180 degree screen surrounding the audience.  

In its theatrical form, an example of the combination of ancient perspective and contrast techniques with contemporary spatialized narratives to bring three dimensional realism to life is the opera \textit{Le Grande Macabre} by Fura dels Baus. Premiered in Brussels in March 2009, the company made use of the correspondence of projected images and physical objects, profiting from the emerging frictions between the projected images and a large scale anthropomorphic figure that mutates in identity and function within the play. The proposal of Fura dels Baus intends to avoid mimesis, and rather than repeat reality, aims at constructing a type of illusory live experience~\cite{fura}.
Several remarkable examples are found in the works of Robert Lepage, and among them, \textit{Needles and Opium}, which narratively relies on an strong engagement with the space, thus the need for a careful representation of location and context within a semi cubic structure that serves as a group of screens that mutate according to the needs of each scene~\cite{pluta}. The use of these types of neutral spatial cavities that serve as immersive screens has grown in popularity during the last decade. In 2012, as part of a marketing campaign for the Sony Playstation products, the agency Marshmallow Laser Feast presented the video mapping of an entire living room using choreographed live puppetry, controlling every element in the scene and performed in front of a non stop recording camera.
The resulting recorded performances offered a high degree of illusion while consolidating the archetypical centralized and predetermined viewpoint, in this case, the viewpoint of the non stop recording camera.

\subsection*{Interactive virtuality}
The advent and popularization of the interactive computing technologies, a variety of spatialized works of art speculate in the direction of producing full body perceptual illusions no longer tied to the spectator's spatial positions and viewpoints predetermined beforehand, but affording contents that interact with the user, whose three dimensional representations are dynamically adjusted to the view point of the user in real time~\cite{svoboda, szel, majum}.

From this point on, a significant part of research from academic and industry related fields, coincided in the aim for developing techniques and devices for visually compositing virtual objects within real environments, with ever increasing interactive faculties for spectators to become co participants. A significant advance was made in this regard by Carolina Cruz Neira, Daniel J. Sandin, and Thomas A. DeFanti at the University of Illinois, Chicago Electronic Visualization Laboratory in 1992. They presented the notion of \textit{CAVE} acronym for \textit{Cave automatic virtual environment}, a semi cubic installation made up of rear high resolution projection screen walls and floor, and a motion capture system that allowed users to interact by physical movements~\cite{cruz, cruz2,prince}. In 1995 a team of researchers from The University of North Carolina at Chapel Hill, including Ramesh Raskar, Greg Welch, and Henry Fuchs presented The office of the future a type of CAVE application not supported on retro projected surfaces but projection on real office surfaces and objects, moreover incorporated real time computer vision techniques to dynamically extract per pixel depth and reflectance information for the visible surfaces in the office including walls, furniture, objects, and people, and then to either project images on the surfaces, render images of the surfaces, or interpret changes in the surfaces~\cite{raskar, raskar3, wye}. 

In 2006 Shaw initiated research into a  fully 360 degree 3D projection system at the UNSW iCinema Research Centre, a centre directed by Shaw and co founded with Dennis Del Favero. AVIE \textit{Advanced Visualisation and Interaction Environment} was first installed in the UNSW Scientia building in 2003 and presented a paradigm that allowed the exploration of 360 degree panoramic projected images, within an architectural framework that correlated the design of the virtual landscape with that of the installation itself, thus simultaneously co activating the virtual representations and the real materials of the projection space.

This installation also introduced a fundamental level of user control, allowing the viewer to control the rotation of a projected window. The design of AVIE follows the traditions of the Panorama, as well as the search for embodied impact of the Baroque Tromp l'oeil fresco painting techniques ~\cite{shaw2, Shaw2003}. One remarkable implementation example is \textit{The Pure Land: Inside the Mogao Grottoes}, a project by Sarah Kenderdine and Jeffrey Shaw working in the University City Honk Kong in 2012. Using digitization data from the Mogao Grottoes in China this installation emphasizes the spatial illusion via pano­ramic immersion and coherent image display techniques~\cite{mogao}. The contents are staged in a 10 meter diameter by 4 meter high cylindrical AVIE theater, while a handheld interface provides interaction with the rendered images, allowing the user to volitionally reveal key parts of the mural paintings on its walls.

These real scale interactive visual display systems allowed the research on the domain of the spatial illusion of reality and helped set the foundations for projector based \textit{Augmented Reality} in the following decades. As defined, Augmented reality aims at building systems that enhances the real world by superimposing computer generated information on top of it, in one or more sensory modalities~\cite{wellner,asselin,billin, caudell}. In its projector based form, Augmented Reality shares the aim to generate perceptual illusion untethered and available to the plain senses, by using synthetic information displayed on real world locations, but distinguishing from mere spatialized motion image in the strong interactive component of the experience~\cite{bimber1, bimber2}.

\section*{Living virtuality}
In the last decades one of the most interesting objects for projector based synthetic augmentation has been the human body. In a collaboration between 1998 and 2002 choreographer Chris Haring and Vienna based director and composer Klaus Obermaier, created two multimedia dance performances, D.A.V.E and Vivisector, that have used the dancer's body as the primary projection surface pre choreographed projections on performers skins that led spectators to believe that the bodies of performers were dramatically reshaped and altered in real time~\cite{ ploe, emph,Dourish2001}. In 2004 together with Ars Electronica Futurelab, Obermaier developed and premiered Apparition, an interactive dance and media performance featuring Desirée Kongerød and Rob Tannion whose movements on stage are captured using an infrared camera and suited with adequate images generated in realtime.\\
At the 8th IEEE International Symposium on Mixed and Augmented Reality ISMAR in 2009, a team of researchers including Greg Welch, Henry Fuchs from The University of North Carolina at Chapel Hill, introduced Shader Lamps Avatars as a new approach for creating a video projected three dimensional avatar using faces of real people. Based on cameras and projectors to capture and map the dynamic motion and appearance of a real person onto a humanoid life sized styrofoam head mounted on an animatronic device, the system delivers a dynamic, real time representation of a person to multiple viewers~\cite{ismar1,ismar2}.
In a similar quest, a collaboration between Japanese media artist Nobumichi Asai, make up artist Hiroto Kuwahara and digital image engineer Paul Lacroix resulted in the Omote project in 2014, using projection mapping techniques to put make up and effects onto a model's face in real time. By using a set of infrared markers painted on the model's face, the team was able to capture and process marker position, estimating face position and orientation, rendering face CG model with animated texture upon, and finally send the image through the projector. This face changing system brings to life a fluid stance of identity in a way similar to the ancient Bian Lian techniques that allow Sichuan Opera's performers to unperceivably change masks in front of spectators.
Further advancing these achievements, The Ichikawa Senoo Laboratory from the University of Tokio aimed in 2015 at overcoming the limited static or quasi static conditions required for video projection mapping. A dynamic projection mapping system was developed, capable of adjusting images in real time onto deforming non rigid surfaces. To this end, the team led by Gaku Narita, Yoshihiro Watanabe, and Masatoshi Ishikawa developed two set of technologies addressing the system's main challenges~\cite{narita, narita2}. First, \textit{Deformable Dot Cluster Marker}, a robust infrared acquisition method for capturing deformations from the surface, even capable of handling large deformation and occlusions. Second, a high speed projector system was developed, \textit{DynaFlash}, capable of displaying 8 bit images in frame rates up to 1,000fps with unprecedented  3 milliseconds latencies. These two innovations mark a breakpoint in the quest for spatially distributed illusion, achieving the perceptual effect as if the projected images were printed onto the target surface. The Ichikawa Senoo Laboratory further demonstrated in 2016 their dynamic projection mapping systems actuating onto a deformed sheet of paper and T shirt on one or multiple targets. Considering the possibility for embedding active or passive markers in clothing and everyday objects~\cite{fibar}, these advancements could represent the novel technical feasibility for most surfaces in the real world to become screens, displaying perceptual illusions coherently and seamlessly~\cite{kott, jasch, borra, landi}. 
\section*{Conclusions}
Video projection mapping has been an important step in the evolution of spatially based perceptual illusionism, and on the base of the evolution of its associated techniques and idiosyncrasies a great variety of research lines have been developed, as well as entertainment and artistic applications. Having mastered linear perspective, color contrast, and vantage point control in order to produce static realistic spatial three dimensional representations, in recent years video projection mapping systems paired with computer vision systems  have acquired the capability of user interaction and of dynamically adjusting the spatial characteristics of the images in order to achieve almost perfect correspondence between projected contents and physical space. Considering video projection mapping as an  intermediate stance in the reality - virtuality continuum~\cite{Milgram1994},  it has been shown to produce at least two types of observable results that occur almost simultaneously~\cite{Pfeiffer2013Multisensory,yun2011}. The first occurs at a sensory level, in which the shape, color and texture properties of the physical surfaces are incorporated by the superimposed images or wane in its benefit. For example the visitors of the Camera degli Sposi are deceived into believing that a two dimensional solid surface such as a brick wall disappears to give way to a different landscape in perfect harmony with the existing architecture. The face in the Omote's models can be suddenly full of make up, and the performers in Obermaier's works may seem to be suddenly in raw flesh.
The second observable phenomena concerns the semantic level of interaction between imaginary contents and physical objects. Depending on the semantic load of each of the interacting elements, the merger between superimposed images and material supports may yield a varying degree of identity transference, as such, physical objects may contrast with projected images or completely assume its identity and its network of semantic associations. In these semantic agreement and opposition relationships, the visitors of the Camera degli Sposi may fall under the impression of being at the residence of a powerful dynasty whose decorated oculus attracts angels from the skies. The models in Omote project can suddenly appear to be skeletons or android humanoids, or it may result unimaginable to think on the bodies in Obermaier's Vivisector as human due to their unprecedented flexibility.
Many steps have been taken in making the real world wane in favor of synthetic superpositions that increasingly become more present, in that every surface acts as a screen and every object contains potential of fluid identity and meaning hosted in the domain of virtual representations. As the production of perceptual illusions appear to be in full effect, new possibilities arise for new entertainment paradigms, spatially distributed learning applications, or massive dataset exploration tools that profit from the embodied cognition that results from spatial interaction with digital contents~\cite{Stahl2006,Stahl13}. Importantly, the relentless idea that fueled the quest for perceptual illusion and pervaded across centuries of artistic and scientific experimentation, seems to gain momentum at the current juncture in history: reality does not satisfy desire.
\section*{Acknowledgments}
Support provided Virtual Sense Systems and Hangar Centre for artistic research and production in Barcelona was greatly appreciated.

\bibliography{augmentingreality2019}

\end{document}